\documentclass[12pt,a4paper]{article}

\usepackage{setspace}
\usepackage{authblk}

\usepackage{float}
\restylefloat{table}

\usepackage{graphicx}	 			
\usepackage{amsmath}
\usepackage{amssymb}
\usepackage{mathrsfs}

\usepackage{xr}

\usepackage{amsthm}
\usepackage{graphicx}	

\usepackage{cancel}

\usepackage{algorithm}
\usepackage{algorithmic}

\usepackage[style=authoryear, giveninits=true, backend=biber, maxcitenames=2, maxbibnames=99, dashed=false, uniquename=false, uniquelist=false, natbib=true]{biblatex}
\DeclareFieldFormat[article]{title}{#1}
\DeclareFieldFormat[article]{volume}{\textbf{#1}} 
\DeclareFieldFormat[article]{number}{\mkbibparens{#1}}
\renewbibmacro{in:}{}
\DeclareNameAlias{sortname}{family-given}

\usepackage{enumitem} 
\usepackage{bm}
\usepackage{booktabs,makecell}
\usepackage{epsfig,amstext,amsfonts,color}
\usepackage{mathtools}
\usepackage{subcaption}
\usepackage[hypertexnames=false]{hyperref}
\hypersetup{
     colorlinks = true,
     linkcolor=blue,
            urlcolor=blue,
            citecolor=blue
}
\usepackage{multirow}
\usepackage{latexsym}
\usepackage{xr}

\DeclareMathOperator*{\argmin}{arg\,min}
\newcommand{\ie}{i.\,e.~}

\newcommand{\eg}{e.\,g.~}

\newcommand{\sph}{\mathbb{S}}
\newcommand{\R}{\mathbb{R}}
\newcommand{\E}{\mathbb{E}}

\newcommand*{\dif}{\mathop{}\!\mathrm{d}}

\newcommand{\btheta}{\bm{\theta}}
\newcommand{\bbeta}{\bm{\beta}}

\newcommand{\hbt}{\bm{\hat{\theta}}}
\newcommand{\hbb}{\bm{\hat{\beta}}}

\newcommand{\hh}{\hat{h}}

\newcommand{\bmu}{\bm{\mu}}
\newcommand{\hbmu}{\bm{\hat{\mu}}}
\newcommand{\tbmu}{\bm{\tilde{\mu}}}
\newcommand{\mbmu}{\mathring{\bmu}}

\newcommand{\cd}{\overset{d}{\to}}

\newcommand{\bal}{\begin{equation}\aligned}
\newcommand{\eal}{\endaligned\end{equation}}

\numberwithin{equation}{section}

\newtheorem{theorem}{Theorem}[section]

\newtheorem{corollary}[theorem]{Corollary}

\newtheorem{lemma}{Lemma}
\newtheorem{remark}{Remark}[section]

\theoremstyle{definition}

\allowdisplaybreaks

\title{A Robust Extrinsic Single-index Model for Spherical Data}

\author[1,*]{Houren Hong}
\author[1]{Janice L. Scealy}
\author[1]{Andrew T. A. Wood}
\author[1]{Yanrong Yang}
\affil[1]{Research School of Finance, Actuarial Studies and Statistics, Australian National University, Canberra ACT 2601, Australia}
\affil[*]{Corresponding author's email: houren.hong@anu.edu.au.~~~~~~~~~~~~~~~~~~~~~~~~~~~~~~~~}

\date{}

\makeatletter
\renewcommand\@biblabel[1]{}
\makeatother

\begin{document}

\maketitle

\begin{abstract} 
\noindent Regression with a spherical response is challenging due to the absence of linear structure, making standard regression models inadequate. Existing methods, mainly parametric, lack the flexibility to capture the complex relationship induced by spherical curvature, while methods based on techniques from Riemannian geometry often suffer from computational difficulties. The non-Euclidean structure further complicates robust estimation, with very limited work addressing this issue, despite the common presence of outliers in directional data.
This article introduces a new semi-parametric approach, the extrinsic single-index model (ESIM) and its robust estimation, to address these limitations. We establish large-sample properties of the proposed estimator with a wide range of loss functions and assess their robustness using the influence function and standardized influence function. Specifically, we focus on the robustness of the exponential squared loss (ESL), demonstrating comparable efficiency and superior robustness over least squares loss under high concentration. We also examine how the tuning parameter for the ESL balances efficiency and robustness, providing guidance on its optimal choice. 
The computational efficiency and robustness of our methods are further illustrated via simulations and applications to geochemical compositional data.

\vspace{3.mm}
\end{abstract}	

\begin{refsection}
\section{Introduction}\label{sec:intro}
Sphere-Euclidean regression models a response on the unit sphere $\sph^{d-1}=\{\bm{y}\in \R^d: \|\bm{y}\| = 1\}$ with $d>2$ using Euclidean predictors and finds various applications in real data, such as marine data, household expenditure compositions, paleomagnetic data and cloud formation data \citep{scealy2011regression,scealy2017directional,scealy2019scaled,paine2020spherical}.
However, existing methods are largely parametric, relying on fully specified link functions, which limits their ability to capture the nonlinear relationships induced by the curvature of the sphere. Moreover, the widespread presence of outliers in real data underscores the need for robust estimation. To date, research has primarily focused on robust estimation of mean direction in non-regression settings \citep{watson1986some,ko1988robustness,he1992robust,ko1993robust,kato2016robust,scealy2021analogues}. 

To enhance model flexibility, we consider the single-index model (SIM). In classical Euclidean settings, the SIM postulates that the effect of predictors $X$ on a real-valued response operates only through a projection of $X$ onto a single index, complemented by an unknown link function. This structure effectively captures potential nonlinear relationships, owing to the nonparametric link function, while retaining interpretability via the coefficients of predictors and mitigating the curse of dimensionality 
\citep{powell1989semiparametric,ichimura1993semiparametric,hardle1993optimal,xia2002adaptive,cui2011efm}. 

Despite extensive research on the SIM for Euclidean responses, these methods are not directly applicable to spherical responses. A unique challenge arises from the non-Euclidean structure, where linear operations are inappropriate, leading to the failure of standard estimation techniques, such as the least squares approach. To address this, existing methods can be broadly classified into two categories: \emph{intrinsic} methods, which directly account for the geometry of the sphere, and \emph{extrinsic} methods, which embed the sphere into a higher-dimensional Euclidean space to apply standard estimation methods, coupled with a projection back onto the manifold.

Among intrinsic methods, \emph{partial intrinsic} methods define and operate on \emph{intrinsic residuals} via the Riemannian logarithmic map and use parallel transport (see \cite{lee2003smooth} for definitions), which allows estimation in the tangent space at the mean direction, a linear subspace of $\R^d$ \citep{jupp1987fitting,zhu2009intrinsic,shi2009intrinsic,yuan2012local,cornea2017regression,lin2019intrinsic,kim2021smoothing}. 
Another class of intrinsic regression models, \emph{Fr\'echet regression}, measures the deviation of the response from the mean direction using geodesic distance, allowing estimation by minimizing this deviation.
The seminal work of \citet{petersen2019frechet} laid the foundation for subsequent extensions \citep{dubey2019frechet, chen2022uniform, bhattacharjee2023single, ghosal2023frechet, qiu2024random}.

Although both of the above intrinsic frameworks respect the non-Euclidean geometry, they typically encounter serious computational challenges. Partial intrinsic methods involve a complex mapping and transport process to compute intrinsic residuals, and Fr\'echet regression models often require gradient descent algorithms to minimize the objective function, which may fail to converge to a global minimum. As a result, computational challenges reduce the practical appeal of intrinsic approaches. We provide numerical evidence for these challenges in Section~\ref{sub:simu:shape}.

In contrast, extrinsic approaches simplify computations by embedding the unit sphere $\sph^{d-1}$ into an ambient space $\R^d$, where most computations are performed in the ambient space using Euclidean distance. These methods offer notable computational advantages. \citet{bhattacharya2012extrinsic} highlight the computational efficiency of the extrinsic mean against the intrinsic. \citet{lin2017extrinsic} propose \emph{extrinsic local regression} as opposed to intrinsic local regression, which exhibits comparable model performance to intrinsic models but with a considerably reduced computational burden. 

However, extrinsic regression models have received relatively less attention in the literature, which motivates our study. In this paper, we propose an Extrinsic Single-index model (ESIM) for sphere-Euclidean regression, combining the flexibility of single-index models with the computational efficiency of extrinsic methods.
Furthermore, since the spherical responses are embedded in the ambient space within the extrinsic framework, standard M-estimation can be applied to the embedded responses, which, in turn, enables robust estimation for the ESIM through a flexible choice of loss functions. In this article, we consider the exponential squared loss function for robust regression. 

Our contributions are threefold. First, we propose the ESIM, which provides sufficient flexibility for the curvature of sphere while preserving interpretability of predictors and maintaining computational efficiency. The asymptotic distributions of our estimators are established. As far as we are aware, this is the first extrinsic semi-parametric regression approach developed for a unit vector response.  
Second, our model allows robust estimation for a wide range of loss functions, such as Huber's loss. We assess its robustness against contaminated samples via the influence function (IF). However, due to the compact and concentrated nature of spherical responses, the IF alone may not sufficiently capture the relative effects of outliers \citep{ko1988robustness}. Thus, we also consider the standardized influence function (SIF), which normalizes the IF with respect to the dispersion of spherical data. To the best of our knowledge, this work is the first to study the SIF in the context of robust spherical regression. Finally, we perform a thorough investigation of robust properties for the exponential squared loss (ESL), particularly in the high-concentration setting. We find that, for moderate-dimensional spherical data, the ESL with a carefully chosen tuning parameter can achieve greater efficiency than the least squares loss. Moreover, we elucidate the trade-off between robustness and efficiency, providing theoretical guidance for the selection of an optimal tuning parameter. Complementing these theoretical findings, extensive numerical studies are performed to validate the statistical performance, computational efficiency and robustness of the ESIM.

This article is organized as follows: Section~\ref{sec:ESIM} introduces the extrinsic single-index model and its M-estimation. In Section~\ref{sec:asym}, asymptotic properties of our estimators are provided and their robustness is studied in Section~\ref{sec:ESL:robust}. 
Tuning parameter choice for the ESL is discussed in Section~\ref{sec:tuning}. In Section~\ref{sec:numerical}, we present numerical studies of the robustness, statistical and computational performance of our model plus a real data application to geochemical compositional data. The Conclusion is given in Section~\ref{sec:conclusion}. Proofs and further numerical results are given in the Supplementary Material.

\section{Model and Estimation Methods}\label{sec:ESIM}
\subsection{The model}
Let $Y\in \sph^{d-1}$, a unit vector, denote a spherical response and suppose $X\in \R^p$ is a $p$-dimensional Euclidean predictor. Define the parameter space
\begin{equation}\label{e:ESIM:Theta}
\Theta := \{\btheta\in \R^{p-1}: \|\btheta\|^2 <1 \} \subset \mathbb{R}^{p-1},
\end{equation}
and introduce a transformation
\begin{equation}\label{e:ESIM:beta}
    \bbeta(\btheta) := \left(\sqrt{1 - \|\btheta\|^2}, \btheta^T \right)^T,
\end{equation}
which maps elements from $\Theta$ to a unit vector with a positive first component. Hereafter, we simplify notation by writing $\bbeta$ for $\bbeta(\btheta)$.

The single-index model assumes $Y$ depends on $X$ only through a linear projection $\bbeta_0^T X$, where $\bbeta_0 = \bbeta(\btheta_0)$ for some $\btheta_0\in \Theta$. To ensure identifiability, we impose constraints $\|\bbeta\|=1$ and $\beta_1>0$, where $\beta_1$ is the first component of $\bbeta$. Such conditions are standard for Euclidean responses \citep{carroll1997generalized,lin2007identifiability,cui2011efm}. Proposition 1 in the supplementary material confirms identifiability for our model. 
However, working directly with $\bbeta$ under unit-norm constraints introduces challenges in both theory (\eg handling Lagrange multipliers) and computation (\eg constrained optimization on manifolds). By parameterizing $\bbeta$ through $\btheta\in \Theta$, we bijectively map the constrained space $\{\bbeta\in \R^{p}: \|\bbeta\| = 1, \beta_1 >0\}$ to the open subset $\Theta\subset \R^{p-1}$, enabling standard M-estimation theory.

Following the extrinsic regression framework, we embed the unit vector $Y$ into a higher-dimensional Euclidean space $\R^{d}$ via an \emph{embedding} $\mathcal{J}:\sph^{d-1} \mapsto \R^d$. The extrinsic single-index model (ESIM) is then formulated as an M-estimation problem
\begin{equation}\label{e:ESIM:theta0}
  \btheta_0 = \argmin_{\btheta \in \Theta}\; \E\left[\psi(\mathcal{J}(Y) - \bmu_{\btheta}(U_{\btheta}))\right],
\end{equation}
where $U_{\btheta}:= X^T \bbeta(\btheta)$ is the projection index, $\psi: \R^d \mapsto \R$ is a known loss function (\eg least squares), and $\bmu_{\btheta}(\cdot)$ is the intermediate link function defined by
\begin{equation}\label{e:ESIM:mu}
  \bmu_{\btheta}(u) = \argmin_{\bmu\in \R^d} \,  \E \left[\psi(\mathcal{J}(Y)-\bmu)|U_{\btheta} = u\right].
\end{equation}
The \emph{population link function} $\bmu_{\btheta_0}(u)$ differs from $\bmu_{\btheta}(u)$ when $\btheta\neq \btheta_0$ and is defined in the ambient space, not necessarily on the sphere. Therefore, by normalizing $\bmu_{\btheta_0}(\cdot)$, we define the \emph{population mean curve} on the sphere 
\begin{equation}\label{e:ESIM:mean}
\tbmu(U_{\btheta_0}) = \mathcal{P}(\bmu_{\btheta_0}(U_{\btheta_0}))=\frac{\bmu_{\btheta_0}(U_{\btheta_0})}{\|\bmu_{\btheta_0}(U_{\btheta_0})\|},
\end{equation}
where the \emph{projection map} $\mathcal{P}(\bm{u})=\bm{u}/\|\bm{u}\|$ with $\bm{u}\in \R^d\setminus\{\bm{0}\}$ maps non-zero vectors to the unit sphere.

\begin{remark}
Throughout this article, we assume spherical response data are unit vectors. Hence, the identity map $\iota: \sph^{d-1}\mapsto \R^d$, where $\iota(Y) = Y$, is a natural choice of embedding. Hereafter, we replace $\mathcal{J}(Y)$ with $Y$.
\end{remark}

There is a broad class of loss functions for robust regression, including the Huber loss and Tukey's biweight loss, satisfying conditions~\ref{ass:ESIM:a1}-\ref{ass:ESIM:a5} in the Appendix. 
We focus on the exponential squared loss (ESL), defined as 
\begin{equation}\label{e:ESL}
\psi_{\text{ESL}} (\bm{u}) = 1 - \exp\{-\|\bm{u}\|^2/\lambda\},
\end{equation}
where $\lambda>0$ is a tuning parameter controlling the trade-off between robustness and efficiency (see Section~\ref{sec:tuning}). As a \emph{redescending} loss function \citep{maronna2019robust}, the ESL is particularly advantageous for spherical responses since its value on outliers decays exponentially with residual magnitude. While its robustness in Euclidean linear regression has been explored by \citet{wang2013robust}, we extend the analysis to the semi-parametric spherical regression in Section~\ref{sec:ESL:robust}.
\begin{remark}
The ESL in~\eqref{e:ESL} is rotationally symmetric in the sense that $\psi_{\text{ESL}} (R\bm{u}) = \psi_{\text{ESL}} (\bm{u})$ for any orthogonal matrix $R$. Nevertheless, it is expected to perform well even under distributions with ellipse-like symmetry, such as the ESAG \citep{paine2020spherical} and SvMF \citep{scealy2019scaled}. This is further supported by our simulation results presented in Section S5.2 of the supplementary material.
\end{remark}

\subsection{Implementation}\label{sub:algo}
The mutual dependence between the parametric component $\btheta$ and the nonparametric component $\bmu_{\btheta}$ complicates the estimation of SIMs. Classical methods typically alternate between estimating $\btheta$ and $\bmu_{\btheta}$ iteratively, resulting in computational complexity. To simplify computation, we adopt the joint estimation approach of \citet{hardle1993optimal}, which estimates both components simultaneously.

Specifically, let $\{X_i, Y_i\}_{i=1}^n$ be an independent and identically distributed sample and denote the projection index $U_{\btheta,i}:= X_i^T \bbeta(\btheta)$, where $\bbeta(\btheta)$ is defined in~\eqref{e:ESIM:beta}. Given $\btheta\in \Theta$, we estimate the link function and its first derivative, $\bmu_{\btheta}$ and $\bmu'_{\btheta}$, via robust local linear regression \citep{fan1994robust,lin2009local}, defined as
\begin{equation}\label{e:ESIM:muhat}
\left(\hbmu_{\btheta, h}(u), \hbmu'_{\btheta,h}(u)\right) = \argmin_{\bm{a},\bm{b}\in \R^d} n^{-1}\sum_{i=1}^n \psi\left(Y_i - \bm{a} - \bm{b}(U_{\btheta,i} - u)\right) K_h(U_{\btheta,i} - u),
\end{equation}
where $K_h(\cdot) = K(\cdot/h)/h$ is a kernel function, and the bandwidth $h$ controls the estimation accuracy of the nonparametric part.
Since $\hbmu_{\btheta, h}$ involves only $\btheta$ and $h$, the simultaneous estimation of both parts is achieved by incorporating it into the criterion function for $\btheta$. In practice, we use the leave-one-out estimator $\hbmu_{\btheta}^{(-i)}$ (see \cite[pp.63-65]{wand1994kernel}), and define the joint estimator as
\begin{equation}\label{e:ESIM:jointloss}
\left(\hbt, \hh\right) =  \argmin_{\btheta \in \Theta;\; h\in(0,\infty)}\; \hat{S}(\btheta, h),\quad \text{where } \hat{S}(\btheta, h) = n^{-1}\sum_{i=1}^n \psi\left(Y_i - \hbmu_{\btheta,h}^{(-i)}(U_{\btheta, i})\right).
\end{equation}
Denote the estimator of $\bbeta$ by $\hat{\bbeta}:=\bbeta(\hbt)$ and write $U_{\btheta_0}:= X^T \bbeta(\btheta_0)$ and $\hat{U}:= X^T \hat{\bbeta}$. Now define the \emph{extrinsic kernel estimate} of $\tbmu(U_{\btheta_0})$ as
\begin{equation}\label{e:ESIM:meanest}
    \check{\bmu}_{\hbt, \hh}(\hat{U}) = \frac{\hbmu_{\hbt,\hh}(\hat{U})}{\| \hbmu_{\hbt,\hh}(\hat{U})\|}.
\end{equation}

Different optimization strategies for solving~\eqref{e:ESIM:jointloss} should be considered depending on the choice of loss function. For the least-squares loss, the nonparametric estimator in~\eqref{e:ESIM:muhat} admits an analytical solution \citep{wand1994kernel}. Substituting it into $\hat{S}(\btheta,h)$ allows direct optimization over $\btheta$ and $h$, which considerably simplifies computation.
However, for loss functions like the ESL in~\eqref{e:ESL}, where no closed-form solution exists for $\hbmu_{\btheta,h}$, the simultaneous estimation must be approached differently. We leverage iteratively reweighted least squares to tackle optimization specific to the ESL. Detailed algorithms for both least squares loss and ESL are provided in Section S4 of the supplementary material. 

\begin{remark}
Apart from advantages in computational efficiency, the joint estimation also facilitates theoretical analysis. By decomposing $\hat{S}(\btheta,h)$, the asymptotic proofs for $\hbt$ and $\hbmu_{\hbt,\hh}$ are simplified, and the asymptotically optimal bandwidth can be derived \citep{hardle1993optimal}. In Section S2.1 of the supplementary material, we extend this decomposition to the M-estimation framework.
\end{remark}

\section{Asymptotic Results}\label{sec:asym}
In this section, we establish the asymptotic distributions of $\hbt$ and $\hbmu_{\hbt,\hh}$. We define the $p\times (p-1)$ Jacobian submatrix
\begin{equation}\label{e:ESIM:jacob}
J(\btheta) = \frac{\partial \bbeta(\btheta)}{\partial \btheta} = 
\begin{bmatrix}
-\frac{\btheta^T}{\sqrt{1- \|\btheta\|^2}}\\
I_{p-1}    
\end{bmatrix}.
\end{equation}
Let $\Sigma_{X|U_{\btheta_0}}=\E[(X-\E[X|U_{\btheta_0}])(X-\E[X|U_{\btheta_0}])^T|U_{\btheta_0}]$ be the conditional covariance matrix of $X$, and define the $(p-1)\times (p-1)$ matrices
\begin{align}
W_0 &= \E\left[\bmu'_{\btheta_0}(U_{\btheta_0})^T F(U_{\btheta_0};\btheta_0) \bmu'_{\btheta_0}(U_{\btheta_0}) J(\btheta_0)^T \Sigma_{X|U_{\btheta_0}} J(\btheta_0)\right],\label{e:ESIM:W0} \\
M_0 &= \E\left[\bmu'_{\btheta_0}(U_{\btheta_0})^T G(U_{\btheta_0};\btheta_0) \bmu'_{\btheta_0}(U_{\btheta_0}) J(\btheta_0)^T \Sigma_{X|U_{\btheta_0}} J(\btheta_0)\right], \label{e:ESIM:M0} 
\end{align}
where $G(U_{\btheta_0};\btheta_0)$ and $F(U_{\btheta_0};\btheta_0)$ are defined in Assumption~\ref{ass:ESIM:a3}. Details of the assumptions are provided in the Appendix. Then, we are ready to establish the following asymptotic distribution for $\hbt$.
\begin{theorem}\label{t:ESIM:thetaAsy}
If Assumption \ref{ass:ESIM:a1}-\ref{ass:ESIM:a5} are satisfied, $W_0$ is invertible and $\btheta_0$ and $\hbt$ are as per \eqref{e:ESIM:theta0} and \eqref{e:ESIM:jointloss}, respectively, then
$\sqrt{n}(\hbt - \btheta_0) \cd N(\bm{0}_{p-1}, W_0^{-1} M_0 W_0^{-1})$.
\end{theorem}

The asymptotic normality of $\hat{\bbeta}:= \bbeta(\hbt)$ is immediate from Theorem~\ref{t:ESIM:thetaAsy} with an application of the multivariate delta method.
\begin{corollary}\label{c:ESL:asy:hbt}
With the same conditions in Theorem~\ref{t:ESIM:thetaAsy}, $\sqrt{n}\left(\hat{\bbeta} - \bbeta_0\right)$ is asymptotically normal with mean $\bm{0}_p$ and singular covariance matrix $J(\btheta_0)W_0^{-1} M_0 W_0^{-1} J(\btheta_0)^T$.
\end{corollary}

\begin{remark} 
The singularity of the asymptotic covariance matrix in Corollary~\ref{c:ESL:asy:hbt} arises from the identifiability constraint $\vert \vert \bbeta_0 \vert \vert =1$ on $\bbeta_0$, rather than the geometric constraints of the response. Since the spherical response is embedded in the ambient space, the asymptotic results of the parametric component aligns closely with that of single-index model in Euclidean settings. This extrinsic framework thus provides a more tractable and computationally efficient alternative than intrinsic methods. 
\end{remark}

Next, we present the asymptotic result for the extrinsic estimate $\check{\bmu}_{\hbt, \hh}$, as defined in~\eqref{e:ESIM:meanest}. This result differs from that of the single-index model in Euclidean settings, because the nonparametric estimates in the ESIM must be projected back to the sphere. Due to the lack of a linear structure, establishing the asymptotic result directly on the sphere is not feasible. Consequently, we conduct the asymptotic analysis at a suitable tangent space on the sphere. 

Recall that $\mathcal{J}:\sph^{d-1}\mapsto \R^d$ denotes the natural embedding. Let $\tilde{M}$ denote the image of this embedding, \ie $\tilde{M}=J(\sph^{d-1})\subset \R^d$. For $\mbmu \in \R^d\setminus\{\bm{0}_d\}$, let $T_{\mathcal{P}(\mbmu)}\tilde{M}$ denote the tangent space of $\tilde{M}$ at $\mathcal{P}(\mbmu)$, where $\mathcal{P}(\mbmu) = \mbmu/\|\mbmu\|$. 
The differential of $\mathcal{P}$ at $\mbmu$ is denoted as $d_{\mbmu}\mathcal{P}:T_{\mbmu}\R^d \mapsto T_{\mathcal{P}(\mbmu)}\tilde{M}$. Theorem~\ref{t:ESIM:asy:NormMu} establishes the asymptotic normality of $\check{\bmu}_{\hbt,\hh}$ at the tangent space.

\begin{theorem}\label{t:ESIM:asy:NormMu}
Let $\nu_2 = \int \nu^2 K(\nu) \dif \nu$ and $\omega_0 = \int K^2(\nu) \dif \nu$ be positive constants.
Define $u_{\btheta_0}=\bm{x}^T \bbeta(\btheta_0)$, $\hat{u} = \bm{x}^T \bbeta(\hbt)$ and $\mbmu_{\btheta_0}(u_{\btheta_0}) = \bmu_{\btheta_0}(u_{\btheta_0}) + \hh^2 \nu_2 \bmu''_{\btheta_0}(u_{\btheta_0})/2$. If Assumption \ref{ass:ESIM:a1}-\ref{ass:ESIM:a5} are satisfied, for all $\bm{x}\in \mathcal{X}$, we have
\begin{equation*}
\sqrt{n\hh} \; d_{\mbmu}\mathcal{P}\left( \hbmu_{\hbt, \hh}(\hat{u}) - \mbmu_{\btheta_0}(u_{\btheta_0})\right) \cd N\left(\bm{0}_{d-1}, \frac{\omega_0}{f_{U_0}(u_{\btheta_0})} B^T F^{-1}_0 G_0 F^{-1}_0 B
\right),
\end{equation*}
where $f_{U_0}(\cdot)$ denotes the marginal density function of $U_{\btheta_0}=X^T \bbeta(\btheta_0)$, $B = B(\mbmu_{\btheta_0}(u_{\btheta_0}))$ is the $d\times (d-1)$ matrix representation of $d_{\mbmu}\mathcal{P}$ with respect to given orthonormal bases of $T_{\mbmu}\R^d$ and $T_{\mathcal{P}(\mbmu)}\tilde{M}$, and $F_0:=F(u_{\btheta_0};\btheta_0)$ and $G_0:=G(u_{\btheta_0}; \btheta_0)$ as defined in Assumption~\ref{ass:ESIM:a3}.
\end{theorem}
\begin{remark}
The asymptotic behavior of the bandwidth $\hh$ is explored in Proposition~3 of the supplementary material, where $\hh\asymp n^{-1/5}$ is able to achieve the asymptotic optimal bandwidth \citep{wand1994kernel}. This rate matches the standard Euclidean setting, implying that, within the extrinsic framework, the geometric structure of the sphere does not alter the bandwidth selection, to leading order.
\end{remark}
\begin{remark}
The matrix $B=B(\mbmu_{\btheta_0}(u_{\btheta_0}))$ in Theorem~\ref{t:ESIM:asy:NormMu} is non-unique and depends on the choice of orthonormal basis for the tangent space $T_{\mathcal{P}(\mbmu)}\tilde{M}$ \citep[p.38]{bhattacharya2012nonparametric}. Specifically, let $\{F_i\}_{i=1}^{d-1}$ be any orthonormal basis of $T_{\mathcal{P}(\mbmu)}\tilde{M}$ and $\{\bm{e}_j\}_{j=1}^d$ denote the standard orthonormal basis of $\R^d$. The differential $d_{\mbmu}\mathcal{P}$ can be expressed in coordinates as $d_{\mbmu}\mathcal{P}(\bm{e}_j)=\sum_{i=1}^{d-1} b_{ji} F_i$, where the coefficients $\{b_{ji}\}_{i=1}^{d-1}$ form the $j$-th row of $B$.
\end{remark}

\section{Robustness under High Concentration}\label{sec:ESL:robust}
This section examines robust properties of $\hbt$, assessed via the influence function (IF). From the perspective of functional derivative, a bounded IF norm indicates that an infinitesimal perturbation at $\bm{z}=(\bm{x},\bm{y})$ has a limited effect on $\hbt$, often referred to as \emph{B-robustness}.
However, in spherical settings, the IF is inherently bounded for continuous estimators (see Corollary~\ref{c:ESL:brobust}), making it an insufficient robustness measure. To address this, we consider the standardized influence function (SIF), introduced by \citet{ko1988robustness}, which normalizes the IF relative to the dispersion of $\hbt$. A bounded SIF then implies \emph{SB-robustness}, providing a more informative assessment, particularly when the responses are highly concentrated around the mean direction $\tbmu(u_{\btheta_0})\in \sph^{d-1}$, a common situation in directional data. 

To formalize high concentration, we consider a semi-parametric class of conditional distributions for $Y\in\sph^{d-1}$ at $u_{\btheta_0}$ of form $f_{Y|U_{\btheta_0}=u_{\btheta_0}} = c_{d,\kappa, f} \; f(\kappa Y^T \tbmu(u_{\btheta_0}))$,
where $c_{d,\kappa, f}$ is a normalization constant, $f$ is the \emph{angular function} and $\kappa>0$ is the concentration parameter. The level of concentration is then jointly determined by $f$ and $\kappa$, where a larger $\kappa$ generally indicates higher concentration, such as in the von Mises distribution. We thus investigate the IF and SIF as $\kappa\to \infty$ and $f\in \mathscr{A}$, where $\mathscr{A}$ is a class of real-valued functions satisfying conditions (D1)-(D3) (see Section S1 of the supplementary material). The class of population joint densities of $(X,Y)$ under different levels of concentration is defined as
\begin{equation}\label{e:ESL:HighConDens}
\begin{split}
\mathscr{F}_{\kappa} = &\left\{F_0: (X,Y)\in \mathcal{X}\times \sph^{d-1}, \tbmu(U_{\btheta_0})\in \sph^{d-1} \right. \\
& \qquad \left. \text{ and } f_{Y|X} = c_{d,k,f} f(\kappa Y^T \tbmu(U_{\btheta_0})) \text{ with }  f\in \mathscr{A}\right\}.
\end{split}
\end{equation}

\begin{remark}
This high-concentration setting closely aligns with the framework of \citet{paindaveine2020inference}. However, our analysis focuses on the influence function as a functional, with no reference to sample size. In \citet{paindaveine2020inference}, the emphasis is different and they consider both $\kappa$ and $n$ tending to infinity simultaneously. 
\end{remark}

Let $F_0\in \mathcal{F}_{\kappa}$ be the joint population density of $(X,Y)$ at a fixed $\kappa$. We define the dispersion matrix of $\hbt$ at $F_0$ as
\begin{equation}\label{e:ESL:dispersion}
\Sigma_{\hbt; F_0} = 
\mathbb{E} \left[\text{IF}(Z; \hbt, F_0) \text{IF}(Z; \hbt, F_0)^T\right] =W_0^{-1} M_0 W_0^{-1},
\end{equation}
where $W_0$ and $M_0$ are defined as in~\eqref{e:ESIM:W0} and \eqref{e:ESIM:M0}, respectively. Assuming $\Sigma_{\hbt; F_0}$ is invertible, the following theorem formulates the IF and SIF of $\hbt$ at a fixed $\kappa$. 

\begin{theorem}\label{t:ESL:IF} 
Let $X$ belong to a compact set $\mathcal{X}\subset \R^p$ and $Y\in\sph^{d-1}\subset \R^{d}$ be a unit random vector for some fixed integer $d > 2$. Denote $F_0\in \mathcal{F}_{\kappa}$ as the joint population density of $(X,Y)$ and $\delta_{\bm{z}}$ as the degenerate distribution on $\mathcal{X} \times \sph^{d-1}$, which places unit probability on a single observation $\bm{z} = (\bm{x}, \bm{y})\in \mathcal{X} \times \sph^{d-1}$. 
Also, let $U_{\btheta_0}=X^T \bbeta(\btheta_0)$ and $u_{\btheta_0}=\bm{x}^T \bbeta(\btheta_0)$.
Then, under Assumptions \ref{ass:ESIM:a1}-\ref{ass:ESIM:a5}, we have
\begin{itemize}
\item[(i)] The influence function of $\hbt$ at $\bm{z}$ is given by 
\begin{equation*}
\text{IF}(\bm{z}; \hbt, F_0) = \psi'\left(\bm{y}-\bmu_{\btheta_0}(u_{\btheta_0})\right)^T \bmu'_{\btheta_0}(u_{\btheta_0}) W_0^{-1} J(\btheta_0)^T \left(\bm{x} - \E\left[X|U_{\btheta_0}=u_{\btheta_0}\right]\right),
\end{equation*}
where $J(\btheta_0)$ is a $p\times (p-1)$ Jacobian submatrix, as defined in~\eqref{e:ESIM:jacob}, and $W_0$ is an invertible $(p-1)\times(p-1)$ matrix, as defined in~\eqref{e:ESIM:W0}.
\item[(ii)] The standardized influence function of $\hbt$ with respect to $\Sigma_{\hbt; F_0}$ is given by
\begin{equation*}
\begin{split}
&\text{SIF}(\bm{z};\hbt, F_0, \Sigma_{\hbt; F_0}) \\
& = \left|\psi'\big(\bm{y}-\bmu_{\btheta_0}(u_{\btheta_0})\big)^T\bmu'_{\btheta_0}(u_{\btheta_0})\right|\\
&\qquad \times 
\left\{ \left(\bm{x} - \E\left[X|U_{\btheta_0}=u_{\btheta_0}\right]\right)^T 
J(\btheta_0) M_0^{-1} 
J(\btheta_0)^T \left(\bm{x} - \E\left[X|U_{\btheta_0}=u_{\btheta_0}\right]\right) \right\}^{1/2}.
\end{split}
\end{equation*} 
\end{itemize}
\end{theorem}

Intuitively, since $Y\in\sph^{d-1}$ is a unit vector, the influence function of $\hbt$ is naturally bounded. Consequently, $\hbt$ remains B-robust even when using a non-robust loss function, such as the least squares loss $\psi_{\text{LS}}(\cdot) = \| \cdot\|^2$. This result is formally stated in the following corollary.

\begin{corollary}\label{c:ESL:brobust}
Under the conditions of Theorem~\ref{t:ESL:IF}, the gross error sensitivity (GES) of $\hbt$ at $F_0$, defined as $\gamma(\hbt, F_0) := \sup_{\bm{z}\in \mathcal{X} \times \sph^{d-1}} \| \text{IF}(\bm{z}; \hbt, F_0)\|$, is bounded.
\begin{align*}
\gamma(\hbt, F_0) & \le \|W_0^{-1} J(\btheta_0)^T\|_2 \times \\
&\quad \sup_{\bm{z}\in \mathcal{X} \times \sph^{d-1}} \|\psi'(\bm{y}-\bmu_{\btheta_0}(u_{\btheta_0}))\| \|\bmu'_{\btheta_0}(u_{\btheta_0})\| \|\bm{x} - \E[X|U_{\btheta_0}=u_{\btheta_0}]\|  < \infty,
\end{align*}
where $\|\cdot\|_2$ denotes the spectral norm of the corresponding matrix.
\end{corollary}

With the IF and SIF of $\hbt$ at a fixed $\kappa$, we are prepared to explore its robustness under high concentration, where the joint population densities $F_0\in\mathcal{F}_{\kappa}$ vary as $\kappa\to \infty$. In particular, we focus on two specific loss functions: the least squares (LS) and the exponential squared loss (ESL), as defined in~\eqref{e:ESL}. The motivation for the ESL is discussed in Section~\ref{sec:ESIM}. Moreover, the tuning parameter $\lambda$ of the ESL plays a crucial role in the high-concentration setting. Since $F_0$ varies with $\kappa$, we consider a varying $\lambda$ by casting it as an \emph{M-estimator of scale}, satisfying
\begin{equation}\label{e:ESL:lambda}
1 - \E\left[\exp\left\{-\frac{\|Y - \bm{\mu}_{\btheta_0}(U_{\btheta_0})\|^2}{\lambda}\right\}\right] = \delta,
\end{equation}  
where $\delta\in (0,1)$ is a fixed constant governing the trade-off between robustness and efficiency. Detailed discussions on the choice of $\delta$ are deferred to Section~\ref{sec:tuning}. With these preparations, we are ready to present the robustness of $\hbt$ under high concentration.

\begin{theorem}\label{t:ESL:GES_SGES}
Let $F_0\in \mathscr{F}_{\kappa}$, where $\mathscr{F}_{\kappa}$ is defined in~\eqref{e:ESL:HighConDens}, and define the score function $g_f:=f'/f$, with $f'$ being the first derivative of the angular function $f$. Denote the estimators $\hbt_{\text{LS}}$ and $\hbt_{\text{ESL}}$ in~\eqref{e:ESIM:jointloss}, corresponding to the least squares and exponential squared loss, respectively, with $\lambda$ for $\hbt_{\text{ESL}}$ satisfying \eqref{e:ESL:lambda}. With Assumption~\ref{ass:ESIM:a1}-\ref{ass:ESIM:a5} and (D1)-(D3) in the supplementary material, we have
\begin{itemize}
\item[(i)] The gross error sensitivity of $\hbt_{\text{LS}}$ and $\hbt_{\text{ESL}}$ at $F_0$ are asymptotically bounded under high concentration. Specifically, as $\kappa g_f(\kappa) \to \infty$, $\gamma(\hbt_{\text{LS}}, F_0) \asymp 1$ and
$\gamma(\hbt_{\text{ESL}}, F_0) \asymp (\kappa g_f(\kappa))^{-1/2}$.
\item[(ii)] Let $\gamma^{\ast}(\hbt, \Sigma_{\hbt, F_0}):=\sup_{\bm{z}\in \mathcal{X}\times \sph^{d-1}} \|\text{SIF}(\bm{z},\hbt, F_0) \|$ be the standardized gross error sensitivity (SGES) of $\hbt$, where the dispersion matrix $\Sigma_{\hbt,F_0}$ is defined in~\eqref{e:ESL:dispersion}.
Then, the SGES of $\hbt_{\text{ESL}}$ is asymptotically bounded, while that of $\hbt_{\text{LS}}$ is not bounded. Specifically, as $\kappa g_f(\kappa) \to \infty$,
\begin{equation*}
\gamma^{\ast}(\hbt_{\text{LS}}, \Sigma_{\hbt_{\text{LS}},F_0}) \ge C_1 \sqrt{\frac{\kappa g_f(\kappa)}{d-1}} \to \infty, 
\end{equation*}
and
\begin{equation*}
\gamma^{\ast}(\hbt_{\text{ESL}}, \Sigma_{\hbt_{\text{ESL}}, F_0}) \le C_2 \sqrt{\frac{c_{\delta} + 4 }{d-1}}
\left(1 + \frac{4}{c_{\delta}}\right)^{\frac{d-1}{4}}  < \infty,
\end{equation*}
where $c_{\delta} = \lim_{\kappa \to \infty} \lambda \kappa g_f(\kappa)$, and $C_1,C_2>0$ are constants independent of $\kappa$.
\end{itemize}
\end{theorem}

\begin{remark}
High concentration is achieved under the condition $\kappa g_f(\kappa) \to \infty$ as $\kappa \to \infty$, as discussed in Section S1 of the supplementary material.
In this article, we use $\kappa \to \infty$ and $\kappa g_f(\kappa) \to \infty$ interchangeably to indicate high concentration. 
\end{remark}

\begin{remark}
Theorem~\ref{t:ESL:GES_SGES} formulates the asymptotic upper bounds for the IF and SIF of $\hbt_{\text{LS}}$ and $\hbt_{\text{ESL}}$ as $\kappa \to \infty$. Part (i) is consistent with Corollary~\ref{c:ESL:brobust}, implying that both estimators maintain B-robust under high concentration. However, part (ii) reveals that $\hbt_{\text{LS}}$ is not SB-robust. Large deviations from the mean direction, relative to the concentration level, results in considerable estimation bias on $\hbt_{\text{LS}}$, as further substantiated by the numerical results in Section~\ref{sub:simu:contaminated}. In contrast, $\hbt_{\text{ESL}}$ is both B-robust and SB-robust due to the redescending properties of the ESL.
\end{remark}

\section{Optimal Tuning Parameter for the ESL}\label{sec:tuning}
The selection of $\lambda$ is crucial for ensuring robustness, particularly in high-concentration settings. The primary challenge, as illustrated in Theorem~\ref{t:ESL:GES_SGES} (ii), is that $\lambda$ must be adjusted accordingly to ensure the limit $c_{\delta}$ exists, since population densities vary with $\kappa$.
The following lemma explicitly formulates the connection between $\delta$ in~\eqref{e:ESL:lambda} and $c_{\delta}$, and thus reducing the problem to choosing an appropriate $\delta \in (0,1)$. The optimal $\lambda$ can then be determined by solving~\eqref{e:ESL:lambda} at different $\kappa$.

\begin{lemma}\label{l:lambda_to_delta}
Suppose the joint distribution for $(X,Y)\in \mathcal{X}\times \sph^{d-1}$ belongs to $\mathscr{F}_{\kappa}$, where $\mathscr{F}_{\kappa}$ is defined in \eqref{e:ESL:HighConDens} for fixed integer $d\ge 3$. If $\lambda$ solves \eqref{e:ESL:lambda} and the limit $c_{\delta} = \lim_{\kappa \to \infty} \lambda \kappa g_f(\kappa)$ exists, we have
\begin{equation}\label{e:ESL:lambda_to_delta}
    \left(\frac{c_{\delta}}{c_{\delta} + 2}\right)^ {(d-1)/2}= 1 - \delta, \qquad \text{as } \kappa \to \infty. 
\end{equation}
\end{lemma}

We now examine the asymptotic relative efficiency (ARE) of the ESL. For clarity, we use subscripts "LS" and "ESL" to indicate quantities associated with the least squares loss and exponential squared loss, respectively. Based on Theorem~\ref{t:ESIM:thetaAsy}, the ARE for the parametric part is then defined as 
\begin{equation}\label{e:ESL:AREpara}
\text{ARE}(\hbt_{\text{ESL}}) := \frac{\text
{Tr}(W_{0,\text{LS}}^{-1}M_{0,\text{LS}}W_{0,\text{LS}}^{-1})}{\text{Tr}(W_{0,\text{ESL}}^{-1}M_{0,\text{ESL}}W_{0,\text{ESL}}^{-1})},
\end{equation}
where $W_0$ and $M_0$ are defined in~\eqref{e:ESIM:W0} and~\eqref{e:ESIM:M0}, respectively. Similarly, leveraging the asymptotic normality of $\hbmu_{\hbt,\hh}$ in Proposition 5 of the supplementary material, the ARE for the nonparametric part at $u$ is given as
\begin{equation}\label{e:ESL:AREnonpara}
\text{ARE}(\hbmu_{\text{ESL}}(u)) := \frac{\text{Tr}\left(F^{-1}_{\text{LS}}(u;\btheta_0) G_{\text{LS}}(u;\btheta_0) F^{-1}_{\text{LS}}(u;\btheta_0)\right)}{\text{Tr}\left(F^{-1}_{\text{ESL}}(u;\btheta_0) G_{\text{ESL}}(u;\btheta_0) F^{-1}_{\text{ESL}}(u;\btheta_0)\right)},
\end{equation}
where $\hbmu(u):=\hbmu_{\hbt,\hh}(u)$ as defined in~\eqref{e:ESIM:muhat}, and $F(u;\btheta_0)$ and $G(u;\btheta_0)$ are $d\times d$ matrices defined in Assumption~\ref{ass:ESIM:a3}. The next result illustrates how the AREs depend on $\delta$. 
\begin{theorem}\label{t:ESL:ARE}
Suppose the joint distribution of $(X,Y)$ belongs to $\mathscr{F}_{\kappa}$, where $\mathscr{F}_{\kappa}$ is defined in~\eqref{e:ESL:HighConDens}. Let $\delta\in (0,1)$ denote the constant which satisfies \eqref{e:ESL:lambda_to_delta}.
Then, as $\kappa \to \infty$, we have
$\text{ARE}(\hbmu_{\text{ESL}}(u)) = R(\delta)^{-1}$ and  
$\text{ARE}(\hbt_{\text{ESL}}) = R(\delta)^{-1}$, 
with positive constants $K(\delta) = 2 /(1-(1-\delta)^{\frac{2}{d-1}})$ and 
\begin{equation}\label{e:ESL:Rdelta}
R(\delta):=  \left(K(\delta) - 2\right)^{-\frac{d-3}{2}}
K(\delta)^{d-1}
\left(K(\delta) + 2\right)^{-\frac{d+1}{2}}
\left(\frac{d-1}{K(\delta)} - 1\right)^{-2}.
\end{equation}
\end{theorem}

\begin{figure}[ht]
\centering
\begin{subfigure}[b]{0.43\textwidth}
    \includegraphics[width=\textwidth]{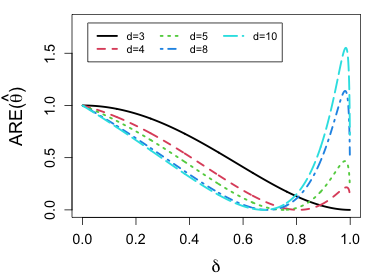}
\end{subfigure}
\begin{subfigure}[b]{0.43\textwidth}
    \includegraphics[width=\textwidth]{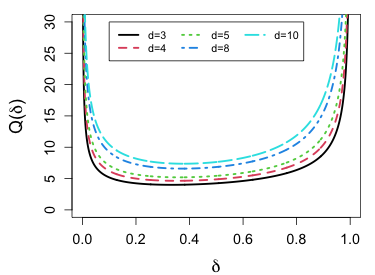}
\end{subfigure}
\caption{Plots for $\text{ARE}(\hbt_{\text{ESL}})$ and $Q(\delta)$  at different response dimensions $d$.} 
\label{fig:ARE_SGES}
\end{figure}

The AREs of both parts are completely determined by $R(\delta)$ as a decreasing function of $R(\delta)$. However, the relationship between $R(\delta)$ and $\delta$ is not straightforward. To better illustrate the role of $\delta$ in the trade-off between robustness and efficiency, we first express the upper bound of the SGES for $\hbt_{\text{ESL}}$ in Theorem~\ref{t:ESL:GES_SGES} (ii) as a function of $\delta$ in the following corollary, illustrating its role in the SB-robustness of $\hbt_{\text{ESL}}$. Figure~\ref{fig:ARE_SGES} then provides graphical illustrations on how $\delta$ influences the trade-off.

\begin{corollary}\label{c:ESL:SGES_delta}
With the conditions of Theorem~\ref{t:ESL:GES_SGES} and $\delta$ satisfying~\eqref{e:ESL:lambda}, the SGES for $\hbt_{\text{ESL}}$ satisfies:
\begin{equation*}
\gamma^{\ast}(\hbt_{\text{ESL}},\Sigma_{\hbt_{\text{ESL}},F_0}) \le C \left(K(\delta) - 2\right)^{-\frac{d-1}{4}}
\left(K(\delta) + 2\right)^{\frac{d+1}{4}}, \quad \text{as } \kappa \to \infty,
\end{equation*} 
where $C>0$ is a constant and $K(\delta) = 2 /(1-(1-\delta)^{\frac{2}{d-1}})$.
\end{corollary}

This upper bound is determined by the convex function 
$$Q(\delta) = \left(K(\delta) - 2\right)^{-\frac{d-1}{4}}\left(K(\delta) + 2\right)^{\frac{d+1}{4}},$$ 
which obtains the global minimum at $\delta_0 = 1 - (1 - d^{-1})^{(d-1)/2}$. 
The right panel of Figure~\ref{fig:ARE_SGES} provides plots of $Q(\delta)$ at different response dimensions $d$, while the left panel illustrates $\text{ARE}(\hbt_{\text{ESL}})$ against $\delta$.

At the left boundary, we observe that $\lim_{\delta\to 0}R(\delta)\to 1$, which implies the ARE for both parts approaches to one, while $Q(\delta)\to \infty$. In this case, the ESL gains efficiency but at the expense of robustness. This loss of robustness is evident from the limiting behavior of the ESL as $\delta\to 0$. By \eqref{e:ESL:lambda_to_delta}, $c_{\delta}=\lim_{\kappa \to \infty} \lambda \kappa g_f(\kappa) \to \infty$, where the ESL approximates to the LS. For instance, considering $\lambda \to \infty$ as $\kappa \to \infty$, the ESL $\psi_{\text{ESL}}(\bm{t}) = 1 - \exp\{- \|\bm{t}\|^2/\lambda \}\approx \|\bm{t}\|^2/\lambda$ for $\bm{t}\in \R^d$ with $\|\bm{t}\|<\infty$. 

Conversely, at the right boundary, we have $\lim_{\delta\to 1}R(\delta)\to \infty$ and $Q(\delta)\to\infty$, leading to a complete loss of both robustness and efficiency for the ESL. In this case, \eqref{e:ESL:lambda_to_delta} yields that $c_{\delta} \to 0$ as $\delta \to 1$, implying $\lambda$ decays too rapidly as $\kappa\to\infty$. As a result, the ESL loses its efficiency entirely as $\psi_{\text{ESL}}(\bm{t}) \to 1$ for $\bm{t}\in \R^d$ with $\|\bm{t}\|<\infty$, potentially resulting in non-uniqueness of $\hbt_{\text{ESL}}$.

Furthermore, when $\delta\in(0,1)$, the monotonicity of $R(\delta)$ with respect to $\delta$ depends on the response dimension $d$, which in turn affects the monotonicity of the ARE. When $d=3$, the ARE decreases monotonically with respect to $\delta$, suggesting a smaller $\delta$ to preserve the efficiency of the ESL. However, for higher-dimensional spherical responses ($d\ge 4$), the ARE is no longer monotonic. Notably, for higher dimensions such as $d=10$, a subtle choice of $\delta$ potentially makes the ESL more efficient than the LS, though this occurs near the region where the ESL loses robustness.

\begin{remark}
Although $\hbt_{\text{ESL}}$ minimizes $Q(\delta)$ at $\delta_0$, this may not be the optimal trade-off between efficiency and robustness. Due to the flat shape of $Q(\delta)$, a range of $\delta\in [0.2,0.6]$, exhibits similar robust properties across dimensions $d$, while a slight increase in $Q(\delta)$ can improve efficiency. Based on empirical observations, we choose $\delta = 0.4$ for stable estimation of $\lambda$ (see Section~\ref{sub:simu:contaminated}).
For more precise selection, a criterion function $L(\delta) = w_1 R(\delta) + w_2 Q (\delta)$ may be constructed, where $w_1,w_2$ are weights reflecting the preference of efficiency and robustness. The optimal $\delta$ is then obtained by minimizing $L(\delta)$, and the corresponding optimal $\lambda$ follows from solving~\eqref{e:ESL:lambda}. 
\end{remark}

\section{Numerical Results}\label{sec:numerical}
\subsection{Estimation with contaminated responses}\label{sub:simu:contaminated}
In this section, we examine statistical performance under contaminated spherical responses, and focus on single-index regression models for fair comparison. These include the extrinsic single-index model with the least-squares loss (LS), the extrinsic single-index model with the exponential squared loss (ESL), the Fr\'echet single-index model (FSIM), and the single-index quantile regression model (SIQR). 

The LS serves as a benchmark, while the ESL provides a robust alternative. The FSIM \citep{bhattacharjee2023single} is its intrinsic counterpart. The SIQR \citep{wu2010single} could, in principle, extend robust single-index models to the extrinsic framework. However, due to challenges in the multivariate extension of quantile regression \citep{konen2022multivariate}, we do not pursue a detailed investigation of SIQR in this work. For this numerical study, we use the $\ell_1$ norm for SIQR, making it a special case of the M-estimator in~\eqref{e:ESIM:jointloss}. Detailed estimation procedures for all models are provided in Section S4 of the supplementary material. 

To evaluate the finite-sample performance, we set $n=200$ and the population index coefficient $\bbeta_0 = (1,-1,1)^T/\sqrt{3}$. The covariates $X_i$, for $i=1, \ldots, n$, are $3$-dimensional vectors following a multivariate normal distribution $N(\bm{0}_3, I_3)$. The mean curve on the sphere is given as $\tbmu(U_i) = (\sqrt{1-v_i^2} \cos(\pi v_i), \sqrt{1-v_i^2} \sin(\pi v_i), v_i)^T$, where $U_i = \bbeta_0^T X_i$ and $v_i = (1+\exp(-U_i))^{-1}$. This is adapted from \citet{petersen2019frechet}, which maps a spiral on the sphere and introduces moderate curvature to each component of $\tbmu(U_i)$ (see Figure S6.9 (b) of the supplementary material).

The responses $Y_i$ for $i=1, \ldots, n$ follow a von Mises-Fisher distribution centered around $\tbmu(U_i)$ with different concentration parameters $\kappa=5,25,50,100,\allowbreak 250$. To introduce contamination, we randomly select a proportion of samples ($\epsilon= 0\%, 10 \%, \allowbreak 20 \%, 30 \%, 40\%$) without replacement, and replace with their orthogonal unit vectors $\tbmu_{\perp}(U_i)$ such that $\tbmu_{\perp}(U_i)^T \tbmu(U_i) = 0$, which highlight the effect of outliers most \citep{ko1993robust}. The orthogonal vectors $\tbmu_{\perp}(U_i)$ are generated by normalizing the cross product between $\tbmu(U_i)$ and a reference vector $\bm{p} = (1,0,0)^T$, ensuring uniqueness by setting $\tbmu(U_i) \neq \bm{p}$ and the first component of $\tbmu_{\perp}(U_i)$ being positive. 
Additionally, we generate 50 pairs of uncontaminated samples, $\{(\tilde{X}_j, \tilde{Y}_j)\}_{j=1}^{50}$, as the test set.

\begin{figure}[ht]
\centering
(a) $\kappa = 5$\\
\begin{subfigure}[b]{0.325\textwidth}
    \includegraphics[width=\textwidth]{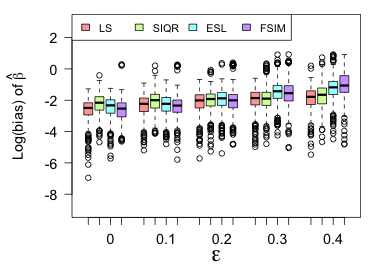}
\end{subfigure}
\begin{subfigure}[b]{0.325\textwidth}
    \includegraphics[width=\textwidth]{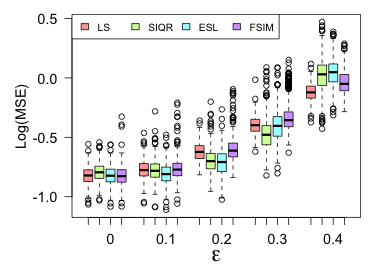}
\end{subfigure}
\begin{subfigure}[b]{0.325\textwidth}
    \includegraphics[width=\textwidth]{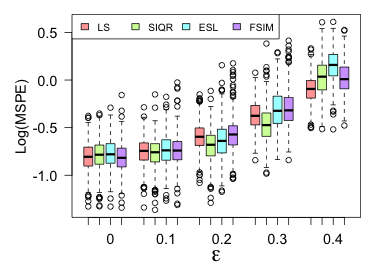}
\end{subfigure}\\ 
(b) $\kappa = 25$\\
\begin{subfigure}[b]{0.325\textwidth}
    \includegraphics[width=\textwidth]{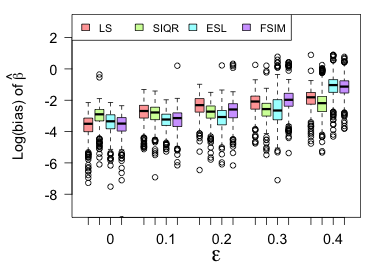}
\end{subfigure}
\begin{subfigure}[b]{0.325\textwidth}
    \includegraphics[width=\textwidth]{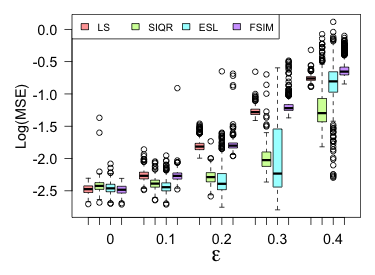}
\end{subfigure}
\begin{subfigure}[b]{0.325\textwidth}
    \includegraphics[width=\textwidth]{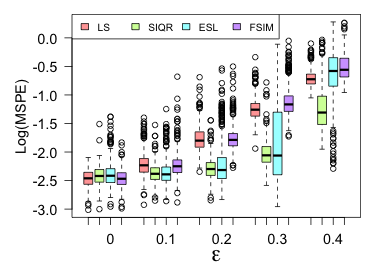}
\end{subfigure}\\ 
(c) $\kappa = 50$\\
\begin{subfigure}[b]{0.325\textwidth}
    \includegraphics[width=\textwidth]{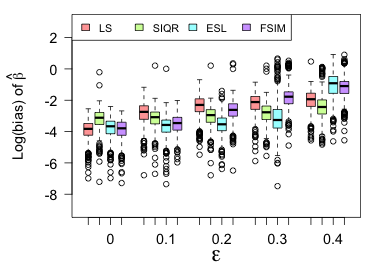}
\end{subfigure}
\begin{subfigure}[b]{0.325\textwidth}
    \includegraphics[width=\textwidth]{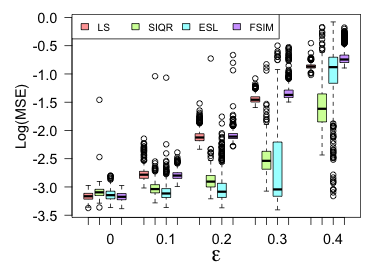}
\end{subfigure}
\begin{subfigure}[b]{0.325\textwidth}
    \includegraphics[width=\textwidth]{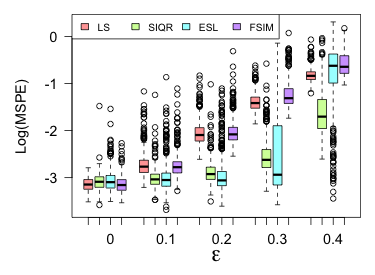}
\end{subfigure}
\caption{Boxplots of the logarithms of bias, MSE and MSPE for contaminated samples, where $\varepsilon$ denotes the proportion of contaminated samples.}
\label{fig:contaminate}
\end{figure}

We repeat this procedure for 500 independent runs. To illustrate model performance, let $\hbb^{(b)}$ denote the estimate of $\bbeta_0$ at the $b$-th run, and define the bias as
\begin{equation}\label{e:bias}
\text{bias}(\hbb^{(b)}) = \arccos(\bbeta_0^T \hbb^{(b)}),
\end{equation}
where $\sup_{\hbb\in \sph^{2}, \hat{\beta}_1>0} \text{bias}(\hbb)=\pi$. Similarly, let $\hat{Y}_i$ and $\hat{Y}_j$ denote the fitted and predicted values, respectively. The mean squared error (MSE) and mean squared prediction error (MSPE) for each repetition are defined as follows, respectively,
\begin{equation}\label{e:mse}
\text{MSE}^{(b)} = \frac{1}{n}\sum_{i=1}^{n}\arccos(Y_i ^T \hat{Y}_i)^2 \quad \text{and}\quad \text{MSPE}^{(b)}=\frac{1}{50}\sum_{j=1}^{50} \arccos(\tilde{Y}_j^T \hat{Y}_j)^2.
\end{equation}
See Figure~\ref{fig:contaminate} for boxplots of the logarithm of bias, MSE and MSPE with different contamination proportions $\epsilon$ and concentration levels, $\kappa=5, 25, 50$. Boxplots for $\hbb$ at other concentration levels are in Section~S6.8 of the supplementary material.

Initially, when the sample is uncontaminated ($\epsilon=0$), the ESL exhibits comparable bias, MSE and MSPE to nonrobust methods, whereas the SIQR shows noticeable bias, which becomes more pronounced as $\kappa$ increases (see $\epsilon = 0$, first column of Figure~\ref{fig:contaminate}). 
Under low concentration ($\kappa =5$, first row of Figure~\ref{fig:contaminate}), robust approaches (SIQR and ESL) fail to outperform nonrobust approaches (LS and FSIM) in terms of bias as $\epsilon$ increases. This failure is attributed to low concentration, where atypical values emerge as inliers, limiting their impact on the estimation. However, the robust approaches achieve slightly lower MSE and MSPE than the nonrobust ones.

As $\kappa$ increases, robust approaches show superior performance under contamination, particularly the ESL. When $\kappa =50$ (third row of Figure~\ref{fig:contaminate}), the ESL achieves the smallest bias among all methods. However, under high contamination ($\epsilon=0.4$), the estimates for the ESL suffer from increased bias and variance due to instability in the iterative estimation of both the model parameters and the tuning parameter $\lambda$. 
When $\epsilon$ is large, outliers heavily influence both components, making the algorithm unstable and difficult to converge. In practice, such extreme contamination is rare. Incorporating a more robust $\hat{\lambda}$ may improve stability in highly contaminated settings. 

\subsection{Shape of mean functions}\label{sub:simu:shape}
The mean function on the sphere is inherently complex due to its curvature, which complicates the estimation of the parametric part. In this section, we examine the statistical performance and computational efficiency of the ESIM across mean functions of varying complexity. 

Specifically, we define the population index as $\bbeta_0 = (1,-1,1,-1,1)^T/\sqrt{5}$. The covariates follow a standard multivariate normal distribution with zero mean and identity covariance matrix. We consider three mean functions, each capturing different aspects of nonlinearity. Specifically, let $U_i = \bbeta_0^T X_i$ and $V_i = \phi(U_i)$ for $i=1,\ldots, n$, where $\phi(\cdot)$ is the cumulative density function of the standard normal distribution. We define $\tbmu_1(U_i) = (\frac{2U_i}{2+U_i^2}, \frac{-2 U_i}{2+U_i^2},\frac{U_i^2 - 2}{U_i^2+2})^T$, 
$\allowdisplaybreaks \tbmu_2(U_i) = ((1 - V_i^2)^{1/2}\cos(\pi V_i), (1 - V_i^2)^{1/2}\sin\left(\pi V_i\right), V_i)^T$ 
and $\tbmu_3(U_i) = \bmu(U_i)/\|\bmu(U_i)\|$ where 
$\bmu(U_i)=(\sin(2\pi U_i), \cos(\pi U_i), \allowbreak2U_i(2+ U_i)^{-1/2})^T$. Figure~S6.9 in the supplementary material demonstrates the shapes of these mean functions. With moderate concentration, \ie $\kappa=20$, we generated the response for each mean curve following the von Mises-Fisher distribution. 

In this numerical study, we generate training sets of size $n=150,300,600$, denoted as $\{(X_i, Y_i)\}_{i=1}^n$, along with a test set of 50 pairs $\{(\tilde{X}_j,\tilde{Y}_j)\}_{j=1}^{50}$. This dataset is applied to the extrinsic single-index model using either the least-square loss (LS) or the exponential squared loss (ESL), as well as to the Fr\'echet single-index model (FSIM) and the extrinsic local regression (Lin), proposed by \citet{lin2017extrinsic}. This procedure is repeated for 500 independent runs, and model performance is assessed in terms of bias, mean squared error (MSE) and mean squared prediction error (MSPE), as defined in~\eqref{e:bias} and~\eqref{e:mse}.

\begin{figure}
\centering
\begin{subfigure}{0.325\textwidth}
    \includegraphics[width=\textwidth]{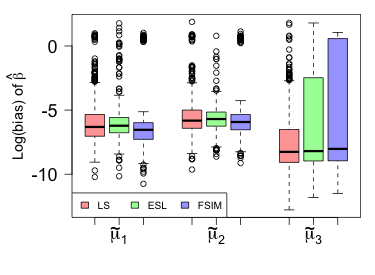}
\end{subfigure}
\begin{subfigure}{0.325\textwidth}
    \includegraphics[width=\textwidth]{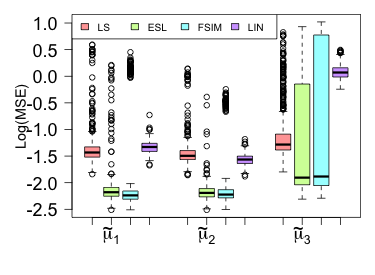}
\end{subfigure}
\begin{subfigure}{0.325\textwidth}
    \includegraphics[width=\textwidth]{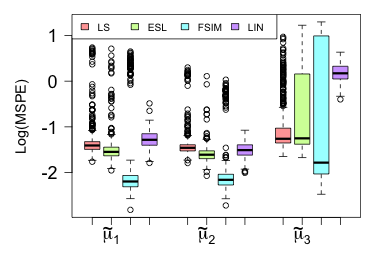}
\end{subfigure}
\caption{Boxplots of the logarithms of bias, MSE and MSPE with different mean functions. (von Mises Fisher distribution with $\kappa = 20$, n=150, replication 500).}
\label{fig:shape:results}
\end{figure}

Figure~\ref{fig:shape:results} summarizes results of size $n=150$ (results for $n=300$ and 600 are provided in Section~S6 of the supplementary material). We first evaluate the bias, where all estimators yield accurate and competitive estimates of $\bbeta_0$. However, when confronted with more complex mean functions, such as $\tbmu_3$, FSIM and ESL are prone to getting trapped at the initial point, leading to poor estimates. This issue can be mitigated with larger sample sizes. Despite these deficiencies, there is no indication that either ESIM or FSIM experiences a larger estimation bias as the complexity of the link function increases.
In terms of MSE and MSPE, FSIM outperforms other methods despite the presence of unusual estimates. Moreover, Lin's performance deteriorates rapidly with increasing shape complexity. 

\begin{figure}
    \centering
    \begin{subfigure}{0.325\textwidth}
        \includegraphics[width=\textwidth]{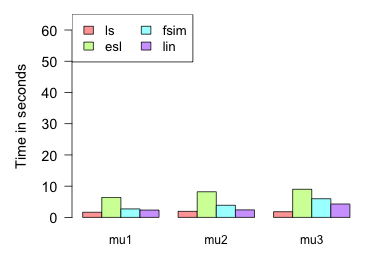}
        \caption{size = 150}
    \end{subfigure}
    \begin{subfigure}{0.325\textwidth}
        \includegraphics[width=\textwidth]{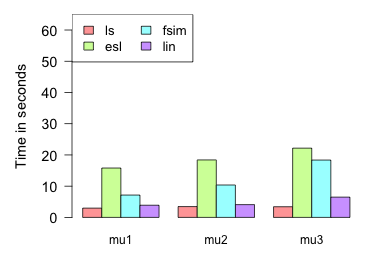}
        \caption{size = 300}
    \end{subfigure}
    \begin{subfigure}{0.325\textwidth}
        \includegraphics[width=\textwidth]{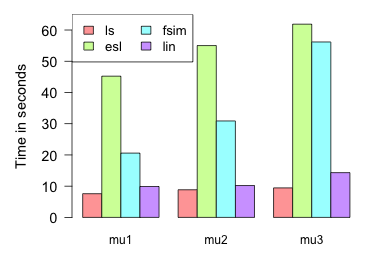}
        \caption{size = 600}
    \end{subfigure}
    \caption{Barplots of the estimation times with different shapes of mean functions and sample sizes. (von Mises Fisher distribution with $\kappa = 20$, replication 500).}
    \label{fig:shape:time}
\end{figure}

To examine the computational advantages of extrinsic approaches, we provide barplots of the sample median of estimation times in Figure~\ref{fig:shape:time}. 
Extrinsic approaches (LS and Lin) consume the least time, with estimation time being insensitive to either the shape of mean functions or the sample size.
The ESL consumes the most time, primarily due to its iterative process (see Section~S4 of the supplementary material). 
In contrast, the estimation time for FSIM considerably increases as the shape complexity or sample size grows.
As discussed in Section~S4, to facilitate the estimation of FSIM, we use extrinsic local regression for mean function estimation, instead of the intrinsic counterpart. Thus, the original estimation method proposed by \citet{bhattacharjee2023single} would consume more time. 

\subsection{Application to soil compositional data}\label{sub:geo}
We analyze soil texture data (weight proportions of sand, silt and clay) from 174 Swedish samples in the Geochemical Mapping of Agricultural Soils (GEMAS) project. 
Soil compositions are converted into spherical responses via the square-root transformation. 
The responses are visualized in Figure~S6.11 in the supplementary material, where some outliers are observed. 

We also consider soil properties from GEMAS as covariates. For categorical data, we consider soil particle size, categorized into four levels: extra large ("ll", 64 samples), large ("l", 65 samples), medium ("m", 44 samples), and small ("s", 1 sample). Due to the imbalance, the "s" category is excluded, leaving 173 samples. The remaining categories are encoded into two indicators, with "l" as the reference level.

For continuous variables, we include climatic factors (mean temperature, MeanTemp, and log-transformed annual precipitation, lAnnPrec) and geochemical factors (log-transformed total organic carbon, ltoc, and soil pH). Log transformations are applied due to skewness (Figure~S6.13). Cation exchange capacity is excluded due to its strong correlation with pH and ltoc (Figure~S6.14).
The final covariates are MeanTemp, lAnnPrec, ltoc, pH and two indicators for soil particle size. Covariates except two indicators are standardized before modelling to cancel the scale effect. 

We apply this data to LS, ESL, FSIM and SIQR, as introduced in Section~S6.1. Although Theorem~\ref{t:ESIM:thetaAsy} provides the asymptotic variances of $\hbb$ for LS and ESL, their practical estimation remains challenging, and those of FSIM and SIQR are unknown. Hence, we estimate standard deviations leveraging the residuals bootstrap, as used in \citet{wu2010single}. However, because a linear structure is unavailable for spherical data, we generate bootstrap samples using rotated residuals \citep{jupp1988residuals}, which defines residuals in the tangent space at a common point, \ie the North pole $Y_0 = (1,0,0)^T$. 

To yield bootstrap samples, rotated residuals are drawn with replacement of size $173$, transported to the tangent space at the fitted values via parallel transport, and mapped back to the sphere via the Riemannian exponential map. Consequently, we fit our model with the bootstrap samples $(X_i, Y_{i}^{\ast})$ and repeat this process 500 rounds. The bootstrap standard deviation of $\hbb$ is then obtained accordingly.

The estimated coefficients are summarized in Table~\ref{tab:geo_soil}, along with the estimated bandwidth $\hat{h}$, MSE and bootstrap standard deviations. All models indicate a significant effect of soil particle size on soil textural composition, except that SIQR considers the level “m” insignificant.  
This is anticipated as a pronounced relationship can be observed in the boxplots for sand-silt-clay texture against soil size (Figure~S6.12). In practice, soil size is a key determinant of soil textural composition. 

Furthermore, the SIQR suggests that soil texture is primarily influenced by particle size, with other predictors being insignificant. In contrast, LS, ESL, and FSIM identify the influence of climatic factors, with MeanTemp significant across all models. Moreover, FSIM also finds lAnnPrec to be significant. Geochemical factors, including ltoc and pH, are insignificant across all models.

\begin{table}[ht]
\centering
\scalebox{0.85}{
\begin{tabular}{l*{8}{c}}
\toprule 
    & ll & m  & MeanTemp & lAnnPrec & ltoc & pH & $\hh$   & MSE \\ \hline
LS  & \textbf{0.5919} & \textbf{-0.7969} & \textbf{0.1171} & -0.0277 & 0.0065  & 0.0083 & 0.0466 &0.0173\\
    & (0.058) & (0.054) & (0.029) &(0.028) &(0.021) & (0.022) &  & \\ 
ESL & \textbf{0.7525}& \textbf{-0.6273} & \textbf{0.1952} & -0.0346& 0.0205  &-0.0207 &0.0215 &0.0172\\
    & (0.080) & (0.149) & (0.039) & (0.019) & (0.017) & (0.020) &  & \\
FSIM& \textbf{0.8586} & \textbf{-0.4906} & \textbf{0.1378} & \textbf{-0.0439} & 0.0246  &-0.0231& 0.0411 &0.0176\\
    & (0.069) & (0.090) & (0.037) & (0.017) & (0.014) & (0.015) &  &\\
SIQR& \textbf{0.9569} &-0.2803  & 0.0424  &-0.0055 &0.0612 & -0.0154 & 0.0675& 0.0213\\
    &  (0.261) & (0.466) & (0.281) & (0.258) & (0.204) & (0.228)     \\
\bottomrule
\end{tabular}
}
\caption{Estimated coefficients of $\bbeta_0$ for soil texture from Sweden, with the bootstrap standard deviation in parentheses.}
\label{tab:geo_soil}
\end{table}

\begin{figure}[H]
    \centering
    \begin{subfigure}[b]{0.24\textwidth}
        \includegraphics[width=\textwidth]{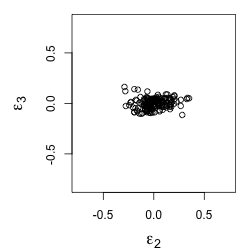}
    \end{subfigure}
    \begin{subfigure}[b]{0.35\textwidth}
        \includegraphics[width=\textwidth]{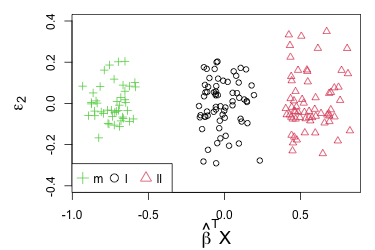}
    \end{subfigure}
    \begin{subfigure}[b]{0.35\textwidth}
        \includegraphics[width=\textwidth]{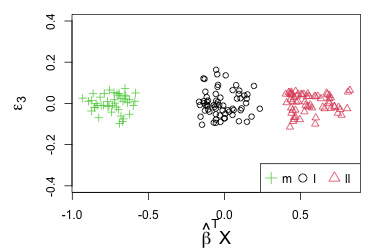}
    \end{subfigure}\\ 
    \begin{subfigure}[b]{0.24\textwidth}
        \includegraphics[width=\textwidth]{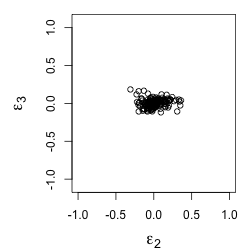}
    \end{subfigure}
    \begin{subfigure}[b]{0.35\textwidth}
        \includegraphics[width=\textwidth]{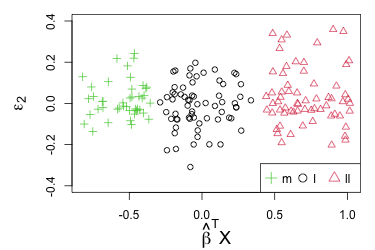}
    \end{subfigure}
    \begin{subfigure}[b]{0.35\textwidth}
        \includegraphics[width=\textwidth]{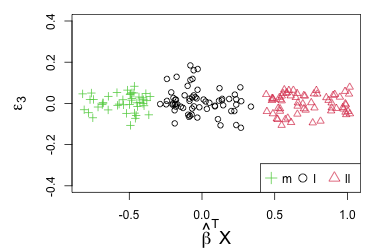}
    \end{subfigure}\\
    \begin{subfigure}[b]{0.24\textwidth}
        \includegraphics[width=\textwidth]{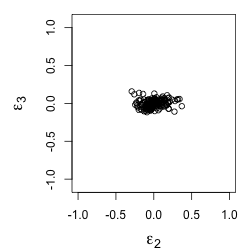}
    \end{subfigure}
    \begin{subfigure}[b]{0.35\textwidth}
        \includegraphics[width=\textwidth]{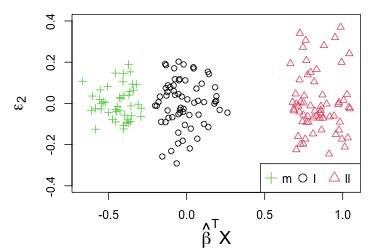}
    \end{subfigure}
    \begin{subfigure}[b]{0.35\textwidth}
        \includegraphics[width=\textwidth]{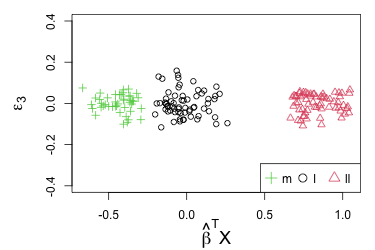}
    \end{subfigure}\\
    \begin{subfigure}[b]{0.24\textwidth}
        \includegraphics[width=\textwidth]{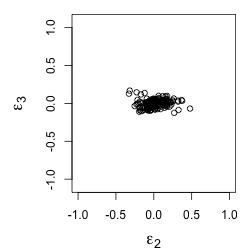}
    \end{subfigure}
    \begin{subfigure}[b]{0.35\textwidth}
        \includegraphics[width=\textwidth]{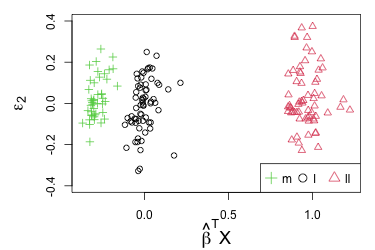}
    \end{subfigure}
    \begin{subfigure}[b]{0.35\textwidth}
        \includegraphics[width=\textwidth]{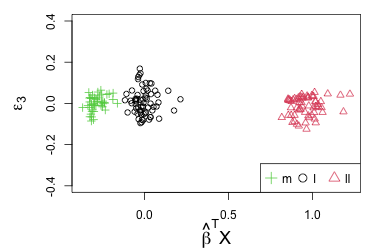}
    \end{subfigure}
    \caption{Scatter plots of the second and third components of the rotated residuals. Each row corresponds to the plots for the LS, ESL, FSIM and SIQR, respectively.}
    \label{fig:sand_res}
\end{figure}

Figure~\ref{fig:sand_res} presents the rotated residuals for all models, with each row corresponding to the plots for the LS, ESL, FSIM and SIQR, respectively. All models indicate that the residuals follow an elliptical distribution, with the second component of residuals showing greater variation compared to the third component (first column of Figure~\ref{fig:sand_res}). The second and third columns of Figure~\ref{fig:sand_res} provide plots of rotated residuals against the fitted index, where a structural pattern is observed. Specifically, the residuals can be grouped by soil size. This is mainly attributed to the discontinuity in $\hbb^T X$. Referring to the estimated coefficients in Table~\ref{tab:geo_soil}, 
the magnitudes of the categorical levels "ll" and "m" dominate those of continuous predictors, leading to this discontinuity. This also suggests that particle size is the most informative predictor for modeling soil textural compositions. 

We also perform a $10$-fold cross-validation to compute MSPE, as defined in~\eqref{e:mse}. The results show that ESL achieves the smallest average MSPE (LS 0.0212, ESL 0.0207, FSIM 0.0292, SIQR 0.0254). Combined with MSE in the last column of Table~\ref{tab:geo_soil}, this indicates that ESL outperforms the others in terms of predictive accuracy.

\section{Conclusion}\label{sec:conclusion}
This article introduces the extrinsic single-index model (ESIM) as a flexible and computationally efficient approach to sphere-Euclidean regression.  As far as we are aware, this is the first extrinsic semi-parametric regression approach developed for a unit vector response. 
By embedding spherical responses in an ambient space, the ESIM enables the application of standard M-estimation, which facilitates the investigation of its asymptotic properties and enabled its robust estimation. 
We assess the robustness of the resulting estimators using the influence function and standardized influence function, addressing the challenges arising from the compactness and concentration of spherical data.
Furthermore, we investigate robustness under high concentration and demonstrate the advantages of exponential squared loss with a carefully chosen tuning parameter, illustrating its improved efficiency over least squares loss in moderate-dimensional spherical settings. We also explore the optimal tuning parameter in the high-concentration scenario from both a theoretical and numerical point of view.

There are several directions for future research. First, the extrinsic framework is applicable to other manifold-valued data, such as shape spaces \citep{lin2017extrinsic}. Thus, it would be valuable to establish unified theoretical foundations for the ESIM in general manifold settings. 
Second, a single index may be insufficient to capture the complex relationships between responses and predictors, which motivates the extension to models with multiple indices. However, designing such models within the extrinsic framework requires careful consideration, since the multidimensional nature of the embedded response affects flexibility and identifiability. Further, the interdependence of responses components makes determining the number of indices more challenging. Lastly, applications to geochemical composition data underscore the need of variable selection methods for spherical regression.

\section*{Appendix}
Below, we summarize the assumptions required to derive theoretical results. Denote the univariate index $U_{\btheta} = X^T \bbeta(\btheta)$ and its realization $u_{\btheta}= \bm{x}^T \bbeta(\btheta)$, where $\bbeta(\cdot)$ is defined in~\eqref{e:ESIM:beta}. 
\begin{enumerate}[label = (A\arabic*), series = fregStrg]
\item \label{ass:ESIM:a1} Assume $X$ belongs to a compact set $\mathcal{X}\subset \R^p$ such that for each $\btheta \in \Theta \subset \R^{p-1}$, where $\Theta$ is as defined in~\eqref{e:ESIM:Theta}, the index $U_{\btheta}$ has a positive marginal density and a continuous second derivative on its support $\mathcal{U}_{\btheta}$. Define $\mathcal{U}=\bigcup_{\btheta\in \Theta} \mathcal{U}_{\btheta}$.
\item \label{ass:ESIM:a2} The link function $\bmu_{\btheta}$ defined in~\eqref{e:ESIM:mu} is Lipschitz continuous: there exists a positive constant $L$ such that $\|\bmu_{\btheta_1}(u_{\btheta_1}) - \bmu_{\btheta_2}(u_{\btheta_2})\| \le L \|\btheta_1 - \btheta_2 \|$, for all $\bm{x}\in \mathcal{X}$ and $\btheta_1, \btheta_2 \in \Theta$.
Further, for all $\btheta\in \Theta$ such that $\btheta \neq \btheta_0$, we assume that 
$P(\bm{x}\in \mathcal{X}: \bmu_{\btheta}(u_{\btheta}) \neq \bmu_{\btheta_0}(u_{\btheta_0}))>0$.
\end{enumerate}
Let $f_{\btheta}(\bm{y}|u)$ be the conditional density of $Y$ given $U_{\btheta}=u$ with respect to unnormalized uniform measure on $\sph^{d-1}$.  
The loss function $\psi$ satisfies the following assumptions.
\begin{enumerate}[label = (A\arabic*), series = fregStrg, start = 3]
\item \label{ass:ESIM:a3} Let $\psi:\R^d \mapsto \R$ be a loss function with $d$-dimensional gradient $\psi'$ and $d\times d$ Hessian matrix $\psi''$. Define $G(u;\btheta)=\E[\psi'(Y-\bmu_{\btheta}(u)) \psi'(Y-\bmu_{\btheta}(u))^T|U_{\btheta}=u]$ and $F(u;\btheta) = \E[\psi''(Y - \bmu_{\btheta}(u))|U_{\btheta}=u]$, which are continuous over $u\in \mathcal{U}$ and $\btheta\in \Theta$. Furthermore, assume both $G(u;\btheta)$ and $F(u;\btheta)$ are invertible and positive definite for $\btheta\in \Theta$.
\item \label{ass:ESIM:a4}
Define $\zeta_{\btheta}(\bm{t}|u) = \E[\psi(Y - \bmu_{\btheta}(u) + \bm{t})|U_{\btheta} =u]$, which has a unique minimizer at the origin, \ie $\zeta_{\btheta}(\bm{0}_d|u)$.
Assume $\zeta_{\btheta}(\bm{t}|u)$ is twice continuously differentiable in $\bm{t}$, with derivatives denoted as $\zeta'_{\btheta}(\bm{t}|u)$ and $\zeta''_{\btheta}(\bm{t}|u)$, respectively. 
For $\tilde{u}\in \mathcal{U}$ in a neighborhood of $u$, we assume that $\zeta_{\btheta}(\bm{t}|\tilde{u})$, $\zeta'_{\btheta}(\bm{t}|\tilde{u})$ and $\zeta''_{\btheta}(\bm{t}|\tilde{u})$ as functions of $\tilde{u}$ are continuous for small $\bm{t}$ with bounded norms.  
\item \label{ass:ESIM:a5}
The conditional density $f_{\btheta}(\bm{y}|u)$ is continuous in $u$. There exist positive constants $\epsilon, \delta>0$ and a positive function $g_{\btheta}(\bm{y}|u)$ such that, for $\tilde{u}$ in a neighborhood of $u$, $\sup_{|\tilde{u} - u|\le \epsilon} f_{\btheta}(\bm{y}|\tilde{u}) \le g_{\btheta}(\bm{y}|u)$. Then, the loss function $\psi$ satisfies
\begin{equation*}
\int \left\| \psi' (\bm{y} - \bmu_{\btheta}(u)) \right\|^{2+\delta} g_{\btheta}(\bm{y}|u) P(\dif \bm{y}) < \infty,
\end{equation*}
and, as $\|\bm{t}\|\to 0$,
\begin{equation*}
\int \left(\psi(\bm{y}-\bm{t}) -  \psi(\bm{y}) - \bm{t}^T \psi'(\bm{y})\right)^2 g_{\btheta}(\bm{y}|u) P(\dif \bm{y}) = o(\| \bm{t}\|^2).
\end{equation*}
\end{enumerate}

Assumption~\ref{ass:ESIM:a1} and~\ref{ass:ESIM:a2} are standard in the literature on single-index models \citep{hardle1993optimal,xia1999single,cui2011efm}. Note that $\mathcal{U}$ is absolutely continuous provided that at least one component of $X$ is absolutely continuous. In Assumption~\ref{ass:ESIM:a2}, we impose a Lipschitz continuity condition to $\bmu_{\btheta}$, ensuring the identifiability of $\btheta_0$ (see Proposition~1 in Section~S2.1 of the supplementary material).

Assumptions~\ref{ass:ESIM:a3}-\ref{ass:ESIM:a5} are commonly encountered in the robust nonparametric regression literature. These conditions are applied to $\zeta_{\btheta}(\cdot|u)$ rather than directly to $\psi$, allowing for robust loss functions with discontinuities, such as the spatial median $\psi_{\text{SM}}(\bm{t}) = \| \bm{t} \|$, which is non-differentiable at the origin. They also imply the following system of (population) estimating equations for $\bmu_{\btheta}$:
\begin{equation}\label{e:ESIM:mufoc}
    \E \left[\psi'(Y-\bmu_{\btheta}(u))|U_{\btheta} = u\right]=\bm{0}_d. \tag{A.1}
\end{equation}
Furthermore, let $\bmu'_{\btheta}(u)\in\R^d$ and $\bmu''_{\btheta}(u)\in\R^d$ denote the first and second derivatives of $\bmu_{\btheta}(u)$, respectively, with respect to $u$. Assumption~\ref{ass:ESIM:a3}-\ref{ass:ESIM:a5} implies that $\bmu'_{\btheta}(u)$ and $\bmu''_{\btheta}(u)$ are continuous differentiable with a bounded norm for all $u\in \mathcal{U}$.

\section*{Funding}
This research was supported by Australian Research Council grant DP2202102232.

\printbibliography
\end{refsection}

\end{document}